\documentclass[aps,twocolumn,nofootinbib,groupedaddress,amsfonts,floatfix]{revtex4-1} 
\usepackage{graphicx,amsmath,amssymb,amstext}
\usepackage{amssymb,amsbsy,amsfonts,amsthm,color}

\usepackage{epsfig}
\usepackage{graphicx}
\usepackage{subfigure}
\usepackage{hyperref}
\usepackage{cancel}

\graphicspath{{Figures/}}

\usepackage{color}
\usepackage[dvipsnames]{xcolor}

\newcommand{\Beq}{\begin{eqnarray}}
\newcommand{\Eeq}{\end{eqnarray}}

\def\lsim{\mathrel {\vcenter {\baselineskip 0pt \kern 0pt \hbox{$<$} \kern 0pt \hbox{$\sim$} }}}

\newcommand{\mpl}{m_{\mbox{\tiny Pl}}}

\def\gsim{\mathrel {\vcenter {\baselineskip 0pt \kern 0pt \hbox{$>$} \kern 0pt \hbox{$\sim$} }}}

\newcommand{\C}{{\mathcal{C}}}

\newcommand{\grchombo}{\mathtt{GRChombo}}

\begin{document}
{\hfill KCL-PH-TH/2018-9}

\title{Gravitational Wave Emission from Collisions of Compact Scalar Solitons}

\author{Thomas Helfer${^\dagger}$, Eugene A. Lim${^\dagger}$, Marcos A.~G.~Garcia${^\ddagger}$, Mustafa A. Amin${^\ddagger}$}
\email{thomashelfer@live.de}
\email{eugene.a.lim@gmail.com}
\email{\mbox{marcos.garcia@rice.edu}}
\email{mustafa.a.amin@gmail.com}
\affiliation{${^\dagger}$Theoretical Particle Physics and Cosmology Group, Physics Department,
Kings College London, Strand, London WC2R 2LS, United Kingdom\\
${^\ddagger}$Department of Physics and Astronomy, Rice University, Houston, Texas 77005-1827, U.S.A.}
\begin{abstract}
We numerically investigate the gravitational waves generated by the head-on collision of equal-mass, self-gravitating, real scalar  field solitons (oscillatons) as a function of their compactness $\C$. We start with solitons that are initially at rest with respect to each other, and show that there exist three different possible outcomes resulting from their collisions: (1) an excited stable oscillaton for low $\C$, (2) a merger and formation of a black-hole for intermediate $\C$, and (3) a pre-merger collapse of both oscillatons into individual black-holes for large $\C$. For (1), the excited, aspherical oscillaton continues to emit gravitational waves. For (2), the total energy in gravitational waves emitted increases with compactness, and possesses a maximum which is greater than that from the merger of a pair of equivalent mass black-holes. The initial amplitudes of the quasi-normal modes in the post-merger ring-down in this case are larger than that of collisions of corresponding mass black-holes -- potentially a key observable to distinguish black-hole mergers from their scalar mimics. For (3), the gravitational wave output is indistinguishable from a similar mass, black-hole--black-hole merger. Based on our results, LIGO may be sensitive to oscillaton collisions from light scalars of mass $10^{-12}\,{\rm eV} \lesssim  m \lesssim  10^{-10}\,{\rm eV}$.   \end{abstract}
\maketitle

\section{Introduction and Results } \label{sect:intro}
The spectacular recent detections of gravitational waves from binary black-hole mergers has heralded a new golden age in gravitational wave physics \cite{Abbott:2016blz,Abbott:2016nmj,Abbott:2017vtc}. Gravitational waves from the merger of compact objects are one of our best resources for probing the strong-field regime of gravity. They also provide us with a probe of the nature of the compact objects themselves.

In addition to black-holes (BH) and neutron stars (NS), the expected quality of the gravitational wave data could allow for the search of exotic compact objects as progenitors in such collisions \cite{Cardoso:2016oxy}. In particular, coherent, self-gravitating, non-topological solitons made of scalar fields are known to have highly compact cores \cite{Friedberg:1986tp, Colpi:1986ye,Seidel:1991zh}. Their collisions may generate observable amounts of gravitational waves and whose waveforms can deviate from those of BH-BH or NS-NS mergers (see in particular \cite{Palenzuela:2006wp,Choptuik:2009ww,Sennett:2017etc,Palenzuela:2017kcg}). 

In this paper, we study the head-on collisions of a class of \emph{real} scalar field solitons called \emph{oscillatons} \cite{1991PhRvL..66.1659S} using $\grchombo$ \cite{Clough:2015sqa} in full general relativity. Unlike boson stars made of complex scalar fields, oscillatons do not have a conserved $U(1)$ charge, but can nevertheless be stable on cosmological time scales \cite{Page:2003rd}. For example, such objects can consist of a spatially localized condensate of an axion field oscillating near the minimum of the potential \cite{Helfer:2016ljl}. Such axion fields are ubiquitous in many high energy physics theories, and are considered to be plausible dark matter candidates (see \cite{Marsh:2015xka} for a review).\footnote{We cannot claim that such compact soliton collisions are likely sources of gravitational waves; an estimate of their population and distribution would be needed, which is beyond the scope of this paper. We hope that the results from this work will motivate such studies further.}
\begin{figure}[t!]
\begin{center}
\includegraphics[width=3.45in]{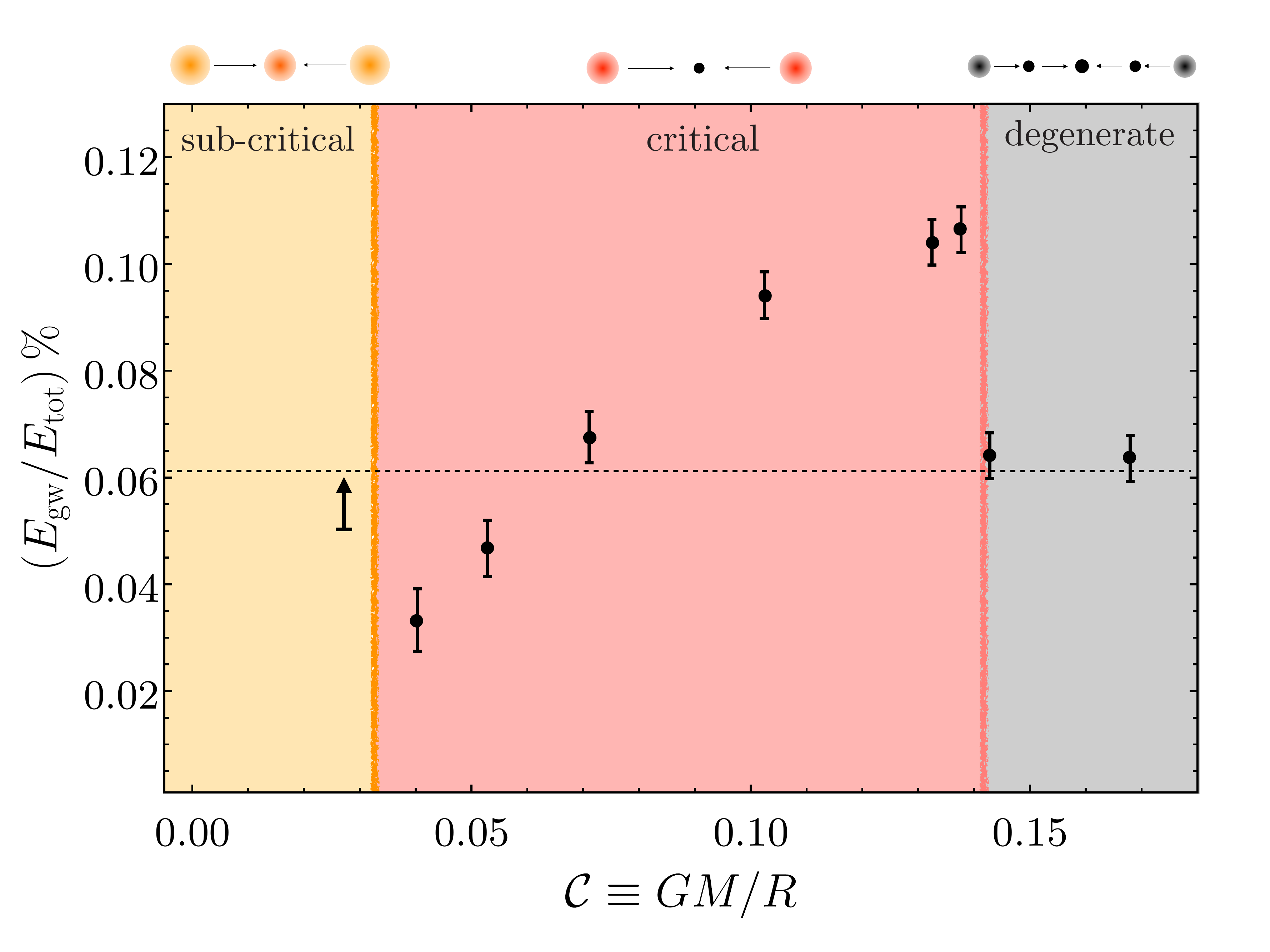}
\vspace{-1.5em} 
\caption{Fraction of initial rest mass energy of the two oscillatons ($E_{\rm tot}$) radiated into gravitational waves ($E_{\rm gw}$) as a function of the initial compactness ($\mathcal{C}$) of each oscillaton. In the subcritical case, oscillatons collide to form a new stable but aspherical, excited oscillaton. In the critical regime, oscillatons collide to yield a black-hole after/during the collision. In the degenerate case, individual oscillatons collapse to black-holes before the collision. Note that in the critical regime (and possibly in the subcritical regime also), the emitted fraction in gravitational waves can exceed that of corresponding mass black-holes ($0.06\%$ dashed line).}
\label{fig:MoneyPlot}
\end{center}
 \end{figure}

Our main result for the gravitational wave output from equal-mass oscillaton collisions as a function of the compactness of the solitons is shown in Fig.~\ref{fig:MoneyPlot}. In particular, the new results are: (1) a jump in the fractional gravitational wave output near a critical compactness value $\C\approx 0.14$, and (2) the fractional gravitational wave output near this $\C$ value exceeds that of corresponding mass black-holes. In order to achieve these results, we constructed \emph{unexcited} oscillaton binaries which possess no spurious additional modes. Such ``clean'' initial conditions allow us to accurately extract the GW production efficiency $E_{\rm gw}/E_{\rm tot}$. Furthermore, we also compute the gravitational waveforms for such collisions to show that the quasi-normal modes are significantly different from equivalent BH-BH collisions during merger and in their ringdown phase, which suggests that they can be distinguished.

\section{Initial Set-Up of Solitons}
We consider a free, massive, real scalar field, which is minimally coupled to gravity with the action\footnote{We use the $-+++$ convention for the metric, and set $\hbar=c=1$. Our Planck mass $\mpl=1/\sqrt{G}$.}
\Beq
S=\int d^4 x\sqrt{-g}\left[\frac{R}{16\pi G}-\frac{1}{2}\partial_\mu\phi\partial^\mu\phi-\frac{1}{2}m^2\phi^2\right]\,,
\Eeq
where $g$ is the determinant of the metric, $R$ is the Ricci scalar, and $m$ is the mass of the real scalar field\footnote{We have ignored possible self-interactions of $\lambda \phi^3$ and higher order terms.} $\phi$. We briefly discuss self-interactions in the Appendix. Conservatively, the results of our paper are expected to apply for solitons made of a sub-dominant axionic dark matter component with the axion decay constant $f\gtrsim m_{\rm Pl}$.  Assuming we have a scenario similar to \cite{Hui:2016ltb}, for $f\gtrsim m_{\rm Pl}$, the total dark matter abundance bound requires the axion to be  unacceptably light ($m\lesssim 10^{-30}\,{\rm eV}$), in conflict with observations~\cite{Marsh:2015xka}. We further discuss this issue and possible solutions in the Appendix.

\begin{figure}[tb]
\begin{center}
{\includegraphics[width=1\columnwidth]{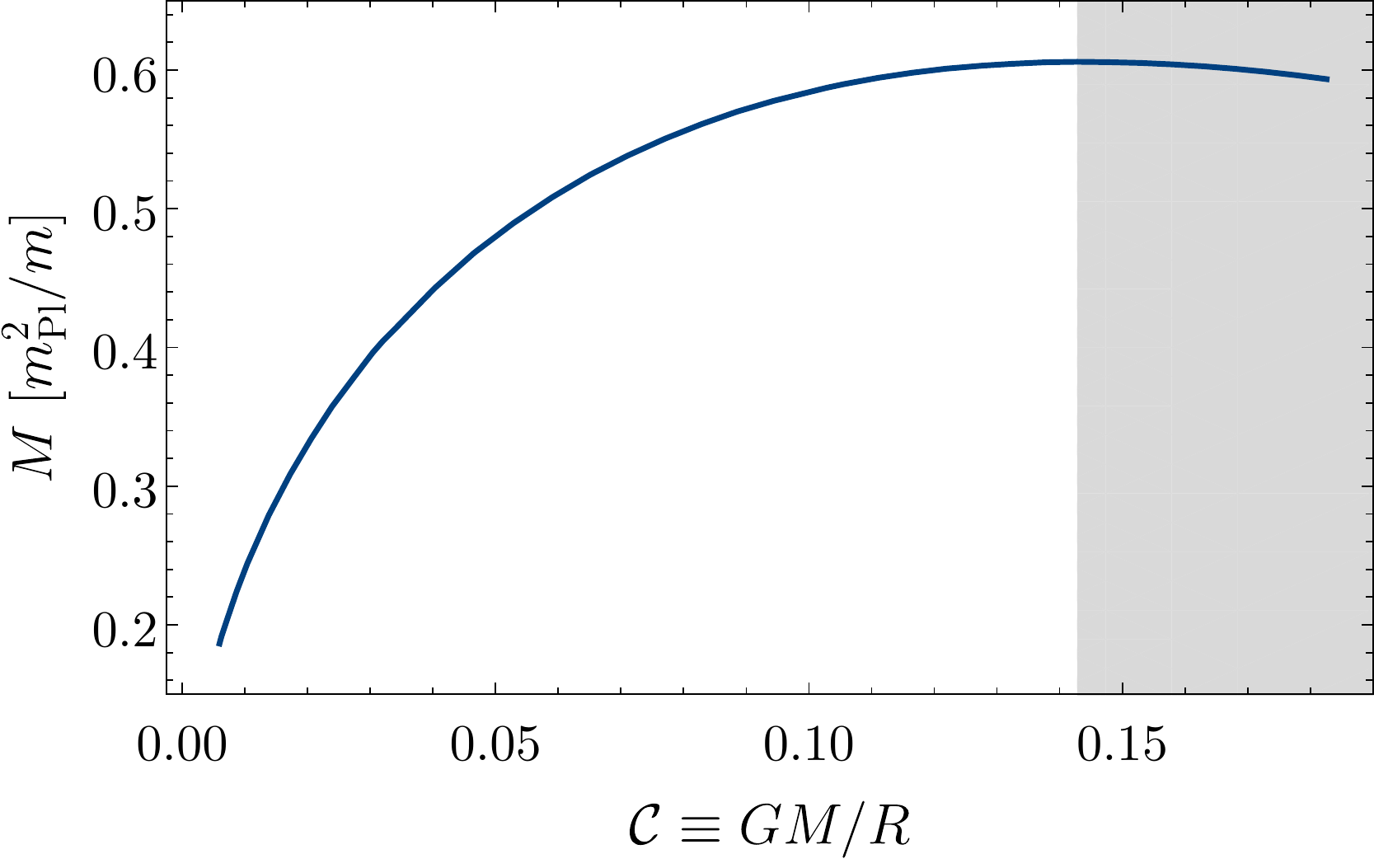}}
\caption{Relationship between the mass $M$ and the compactness $\C$ of the oscillaton. Note that for $\C \gtrsim 0.14$ ($M \approx 0.605\,\mpl^2/m$) oscillatons become unstable under perturbations (grayed area).} 
 \label{fig:MvsC}
\end{center}
\end{figure}

This theory contains a single parameter family of lo\-ca\-li\-zed, solitonic solutions called oscillatons (once the mass $m$ is scaled out). We choose to parametrize our solutions in terms of their \emph{compactness}, which we define as
\begin{equation}
\C \equiv  \frac{GM}{R}, \label{eqn:compactness}
\end{equation}
where $M$ is the Arnowitt-Deser-Misner (ADM) mass, and $R$ is the effective radius of the oscillaton which encompasses $95\%$ of its mass. The maximum mass of the oscillaton $M_{\rm max}\approx0.605 \mpl^2/m$ occurs when $\C\approx0.14$. When $\C<0.14$, the oscillatons are stable against perturbations. For $\C>0.14$, they are unstable with respect to perturbations \cite{Alcubierre:2003sx} (Fig. \ref{fig:MvsC}).

To ensure that these results are qualitatively and quantitatively robust, we implemented several steps such that the initial conditions for these oscillatons are in their unexcited ``ground'' state. We refer the reader to the Appendix for details of this construction,   and other nu\-me\-ri\-cal convergence tests.

We set up two equal $\C$ (and hence equal mass) oscillatons at a distance of $60 m^{-1}$, both of which are initially at rest, and explore the end-state of the collision and gravitational wave signature as a function of $\C$. These oscillatons can also have a relative phase $0<\Delta \theta <\pi$ between their respective oscillations. Oscillatons are considered ``in-phase'' when $\Delta \theta=0$; this is the scenario we focus on in this paper. ``Out-of-phase'' $\Delta \theta \neq 0$ oscillatons exhibit additional \emph{repulsive} force that, at sufficiently large phase differences, prevents a merger from occurring. We will leave the results of out-of-phase initial conditions to a future publication.

\begin{figure*}[t!]
\begin{center}
\includegraphics[width=6.85in]{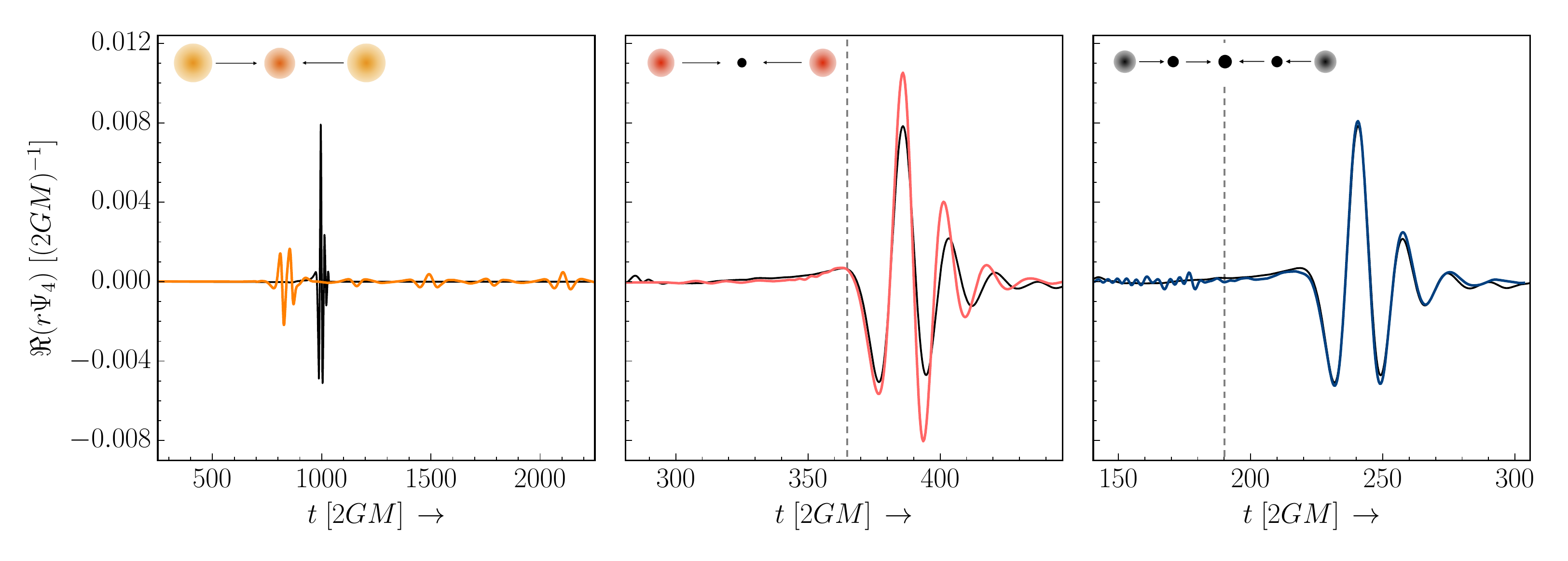}
\caption{The panel shows the numerically evaluated gravitational wave waveforms (the dominant quadrupole mode: $l=2$, $m=0$ is shown) for typical subcritical ($\C=0.03$), critical ($\C=0.10$) and degenerate ($\C=0.15
$) collisions. For comparison, the waveform for the corresponding mass black-hole collision is also shown in black. Note that in the sub-critical case (left panel), the resulting excited oscillaton continues to emit gravitational waves. In the critical case (middle panel), the waveform is qualitatively similar to a BH-BH merger, but importantly, the post-merger QNM amplitude  is greater for the oscillaton merger as they are less ``rigid''. Since there is some mass loss during the merger, the final mass is less than that of the equivalent BH-BH merger, leading to a slightly shorter QNM period (as observed). Finally, the degenerate case is almost indistinguishable from a BH-BH collision (right panel). The vertical dashed line indicates time of BH formation during the merger in the critical case, and pre-merger in the degenerate case. Note that $r\Psi_4\sim r \ddot{h}$, and $t=0$ is associated with the time when the objects are $\approx60m^{-1}$ apart. Movie links for the time evolution of the gravitational wave signal and the energy density $\rho$ are available for the \underline{\color{blue} \href{https://youtu.be/ri2jkxA-a-o}{subcritical}}, \underline{\color{blue} \href{https://youtu.be/FHYvFCSwcaY}{critical}} and \underline{\color{blue} \href{https://youtu.be/R6Hz2Q4FbIU}{degenerate}} mergers~\cite{VidSubc,VidCrit,VidDeg}.}
\label{fig:Waveforms}
\end{center}
 \end{figure*}

\section{Gravitational Waves from Collisions}

We find that there are three possible outcomes of head-on collisions of equal mass solitons depending on the initial $\C$ of the oscillatons.
\subsection{Subcritical Case  $\C\lesssim 0.04$}
\begin{figure}[tb]
\begin{center}
{\includegraphics[width=1\columnwidth]{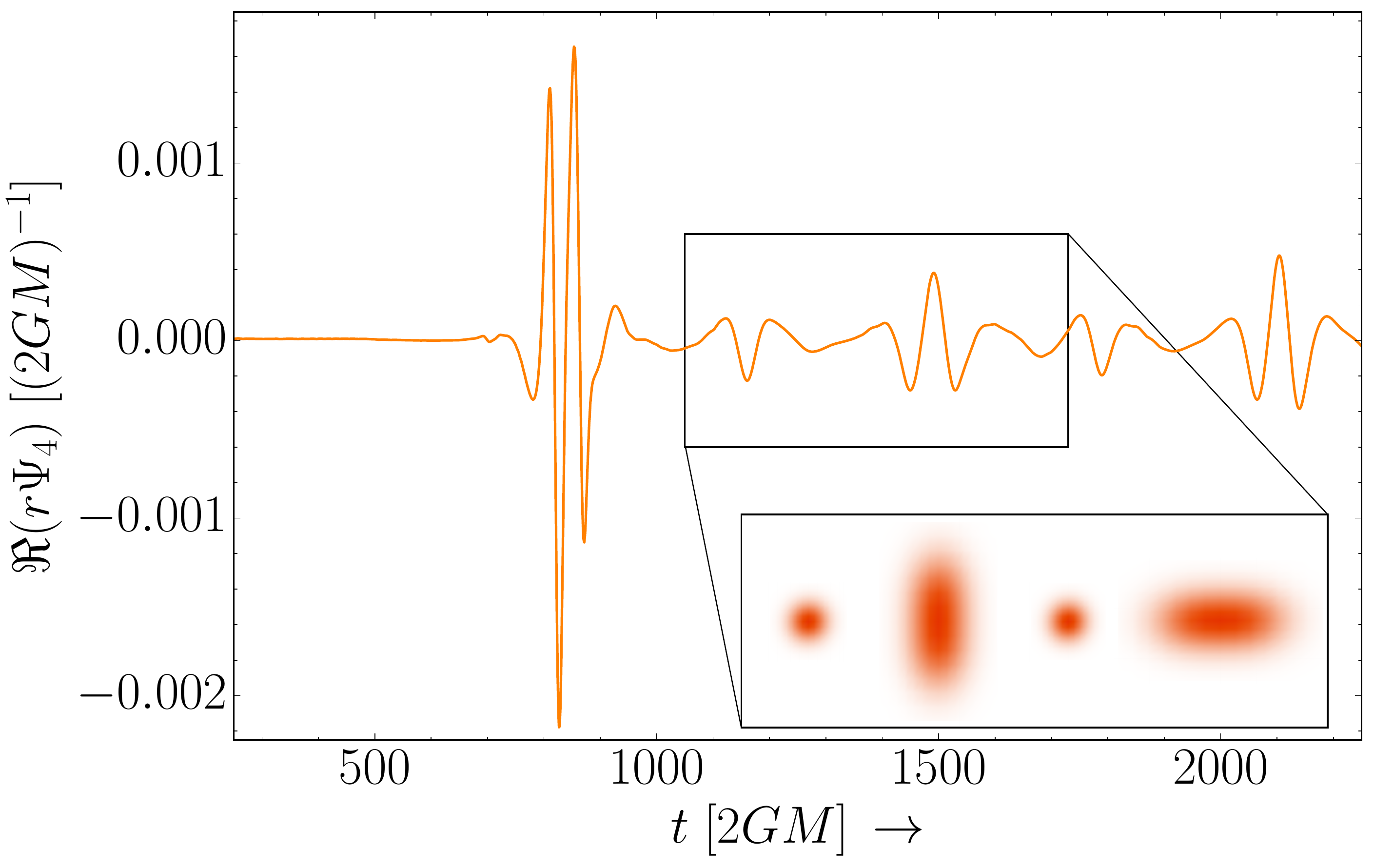}}
\caption{Numerically evaluated gravitational wave waveform for a typical subcritical ($\C=0.03$) collision, demonstrating bursts of repeating gravitational waveforms. Inset shows the $``+"$ pattern of oscillations of the density of the perturbed final state.} 
 \label{fig:WaveForms_sub}
\end{center}
\end{figure}

Collisions of these less compact oscillatons form another more massive and quasi-stable, but aspherical and excited oscillaton. The merger proceeds via multiple stages.  As we have shown in Fig.~\ref{fig:WaveForms_sub}, the oscillatons collide and initially form a very perturbed oscillaton, whose density oscillates in a $``+"$ pattern (i.e.~periodically becomes elongated along two perpendicular axes).  A significant amount of mass is lost during the initial collision. This mass loss can be inferred from the fact that $M (\C=0.03)=0.41\mpl^2/m > 0.5M_{\rm max}$ -- this is consistent with the results first obtained in \cite{Brito:2015yfh}. Without any mass loss, the final oscillaton in this case would have been unstable, which is not seen in our simulations. The oscillaton continues to radiate scalar waves and, notably, a long-time-scale continued emission of gravitational waves \cite{Cardoso:2016oxy,Hanna:2016uhs}. 

From Fig.~\ref{fig:WaveForms_sub} and the first panel in Fig.~\ref{fig:Waveforms}, we see repeated alternating max/min bursts of gravitational waves coinciding with the maximum deformation of the perturbed oscillaton perpendicular-to/along the axis of collision. While the lack of computational resources prevented us from evolving this end state further in time, we expect that the continued emission of both scalar waves and GW will eventually sphericalize the oscillaton. This so-called ``super-emitter'' \cite{Hanna:2016uhs} will eventually emit more total GW energy than the corresponding equal-mass BH-BH merger. We have only found the lower bound on this GW energy output numerically.

\subsection{Critical Case $0.04 \lesssim \C \lesssim 0.14$}

These more massive and compact oscillatons collide to form a black-hole surrounded by a thin scalar field ``wig". There is a slight mass loss during the collision, but the majority ($\approx 90\%$) of the initial mass remains in the final black-hole state. The total GW energy emitted by this merger monotonically scales with $\C$ in this regime. However, interestingly, for oscillatons with $\C > 0.06$, the fraction of the emitted gravitational wave energy\footnote{This energy is computed by integrating over time the total power given by
\begin{equation}
\frac{dE_{\mathrm{gw}}}{dt} = \lim_{r\rightarrow \infty} \frac{r^2}{16\pi} \oint\left|\int_{-\infty}^{t}\Psi_4(r)~dt^{\prime}\right|^2 d\Omega,
\end{equation}
where $\Psi_4$ is the Newman-Penrose scalar. For our simulations the extraction radius $r=60m^{-1}$. Moreover, $E_{\rm tot}=2M$, i.e.~total initial ADM mass of the oscillatons.
}
to the total initial energy, $E_{\rm gw}/E_{\rm tot}$, is \emph{greater} than that from an equivalent head-on merger of a pair of equal mass black-holes ($E_{\rm gw}/E_{\rm tot} = 0.06\%$). The maximum gravitational wave energy emitted $E_{\rm gw}/E_{\rm tot}\approx 0.11\%$ occurs at $\C\approx 0.14$, the boundary where the individual oscillatons themselves become unstable.

A typical waveform of the merger from this region is shown in the middle panel of Fig.~\ref{fig:Waveforms}. Black-hole formation occurs rapidly after the initial merger. For less compact oscillatons, not surprisingly, the scalar dynamics during merger will lead to different GW waveforms that distinguishes it from that of a BH-BH merger \cite{Cardoso:2016oxy, Palenzuela:2017kcg}. Crucially however, even for very compact oscillatons where BH formation is very rapid, the waveform differs from that of a BH-BH collision even in the post-merger ringdown stages.  The quasi-normal mode (QNM) frequency during the ring down is close to that of a BH-BH merger (as expected) with a shorter period due to mass loss during merger. Importantly, the initial amplitude of the QNM is different. In particular, we find that for $\C \gtrsim 0.06$ the initial QNM amplitude is \emph{larger} than that of an equal mass BH-BH merger leading to the aforementioned higher output in total GW emission (see Fig. \ref{fig:MoneyPlot}).  

We believe that this higher initial amplitude for the QNM is due to the fact that oscillatons are ``less rigid'' than black-holes, and hence easier to excite during the initial merger phase. 

Interestingly, in \cite{Cardoso:2016oxy}, the authors argue instead that collisions of more massive boson stars will lead to a more rapid collapse into BHs and hence to a smaller deviation from a BH-BH merger. Our results here show that the deviation is \emph{more} significant for oscillatons, allowing us to directly test for such non-BH merger scenarios.\footnote{ Furthermore we note that while boson star mergers can be qualitatively similar to our oscillaton mergers, there are differences. For example, a collision between a boson star and anti-boson star can lead to annihilation, with a dispersal of most of the field to infinity~\cite{Bezares:2017mzk}. Analogs of boson star/anti-boson star pairs are not present in the oscillaton merger case. Note that an initial phase difference between pre-merger oscillatons cannot mimic these configurations. }

\subsection{Degenerate Case $\C > 0.14$}
Oscillatons with $\C > 0.14$ are inherently unstable to perturbations. We find that as they fall towards one another, mutual tidal forces generate sufficient perturbations to cause the oscillatons to collapse into a pair of BHs before they collide.  The new BHs (with a thin wig) then collide and merge as in the standard BH-BH  case to form a larger black-hole. The waveform (see rightmost panel of Fig. \ref{fig:Waveforms}) and the fraction of energy in gravitational waves is indistinguishable from the BH case and remains constant as we continue to increase the compactness (see Fig. \ref{fig:MoneyPlot}). This expected behavior in the degenerate case makes for a strikingly steep transition in the emitted gravitational wave energy from the critical to the degenerate regime (near $\C\approx 0.14$, see Fig. \ref{fig:MoneyPlot}).

\section{Discussion and Future Directions}

Our main results can be seen in Fig.~\ref{fig:MoneyPlot} and Fig.~\ref{fig:Waveforms}.  Through detailed calculations using the full power of numerical GR, we showed that oscillaton head-on mergers have distinctly different GW signatures than that of their corresponding equal mass BH-BH counterparts. We found three different outcomes of collisions depending on the initial compactness of the oscillatons: formation of excited oscillatons (sub-critical), formation of a black-hole after collision (critical) and formation of black-holes before collision (degenerate) due to tidal forces. 

In terms of gravitational waves, the subcritical merger results in a potentially long lived source of gravitational waves. The gravitational waveform is qualitatively different from the black hole merger case with multiple post merger pulses. For the degenerate case, the dynamics and gravitational wave signatures are very similar to that of corresponding mass black-holes.

Most interestingly, for critical mergers where the final state is a BH, the post-merger QNM mode has a significantly larger amplitude than that of an equivalent BH-BH merger (for sufficiently compact oscillatons). We believe that this is due to the fact that oscillatons are less rigid and easier to excite than BHs. This raises the possibility that, without inspiral GW information, compact oscillatons mergers may mimic BH mergers of a larger mass, though QNM frequency information will allow us to break this degeneracy. Conversely with inspiral information, this may provide a distinct GW signature for the detection of such exotic compact objects.   If these results carry through to inspiral mergers, the ratio of the GW amplitude during the inspiral phase and the ringdown phase will be a strong indicator of exotic mergers. While these are plausible arguments, more work is needed in the inspiral case to make a convincing argument regarding observationally distinguishing BH-BH mergers from OS-OS ones.

For the cases checked (in head-on collisions), we found that the fraction of energy density in gravitational waves is relatively independent of the initial separation of the solitons (within numerical error, we confirmed this for separations of $40m^{-1}$, $55m^{-1}$ and $65m^{-1}$ in the critical and degenerate regimes). The critical-degenerate boundary in Fig. \ref{fig:MoneyPlot} is similarly robust, suggestive of some novel criticality in terms of the dynamics and the gravitational wave output near $\C\approx 0.14$, which is worth investigating in detail. Further investigation of this criticality by scanning through different initial velocities, relative phases and a larger variance in distances is needed. 

Assuming that our oscillatons are stellar mass so that their QNM frequencies fall within LIGO range, this allows us to probe light oscillaton masses of $10^{-12}\,{\rm eV} \lesssim  m \lesssim  10^{-10}\,{\rm eV}$. On the other hand, interactions of free scalar fields with rotating black holes can cause a superradiance instability, robbing the blackholes of their spin -- LIGO (LISA) observations of stellar mass (supermassive) spinning black holes can potentially rule out $10^{-13} \lesssim  m \lesssim  10^{-12}$ eV  ($10^{-19} \lesssim  m \lesssim  10^{-17}$ eV) \cite{Brito:2017zvb}.

In conclusion, we have found that in head-on collisions, compact scalar field solitons can be louder in gravitational waves than their black-hole counterparts. Moreover, a new critical transition in the GW amplitude is seen at $\C\approx 0.14$.  It will be interesting to see if these results are replicated in the inspiral case (e.g.~\cite{Palenzuela:2017kcg}), which we are currently investigating. \\

\acknowledgments
We would like to thank Ricardo Becerril for the use of his initial condition code for oscillatons, and acknowledge useful conversations with Katy Clough, Vitor Cardoso, James Cook, William East, Scott Hughes and Matt Johnson. TH and EL are supported by STFC AGP grant ST/P000606/1. MA and MG are supported by a US.~Dept.~of Energy grant: DE-SC0018216.
We would also like to thank the GRChombo team
(http://www.grchombo.org/) and the
COSMOS team at DAMTP, Cambridge University for
their ongoing technical support. Numerical simulations
were performed on the COSMOS supercomputer, funded by DIRAC/BIS, on BSC Marenostrum IV via PRACE grant Tier-0 PPFPWG, by the Supercomputing Centre of Galicia and La Palma Astrophysics Centre via BSC/RES grants AECT-2017-2-0011 and AECT-2017-3-0009 and  on SurfSara Cartesius under Tier-1 PRACE grant FI-2017-1-0042. This work was done in part at the Aspen Center for Physics, which is supported by National Science Foundation grant PHY-106629. Some simulation results are analyzed
using the visualization toolkit YT \cite{0067-0049-192-1-9}.

\appendix

 \begin{figure}[tb]
\begin{center}
{\includegraphics[width=1.0\columnwidth]{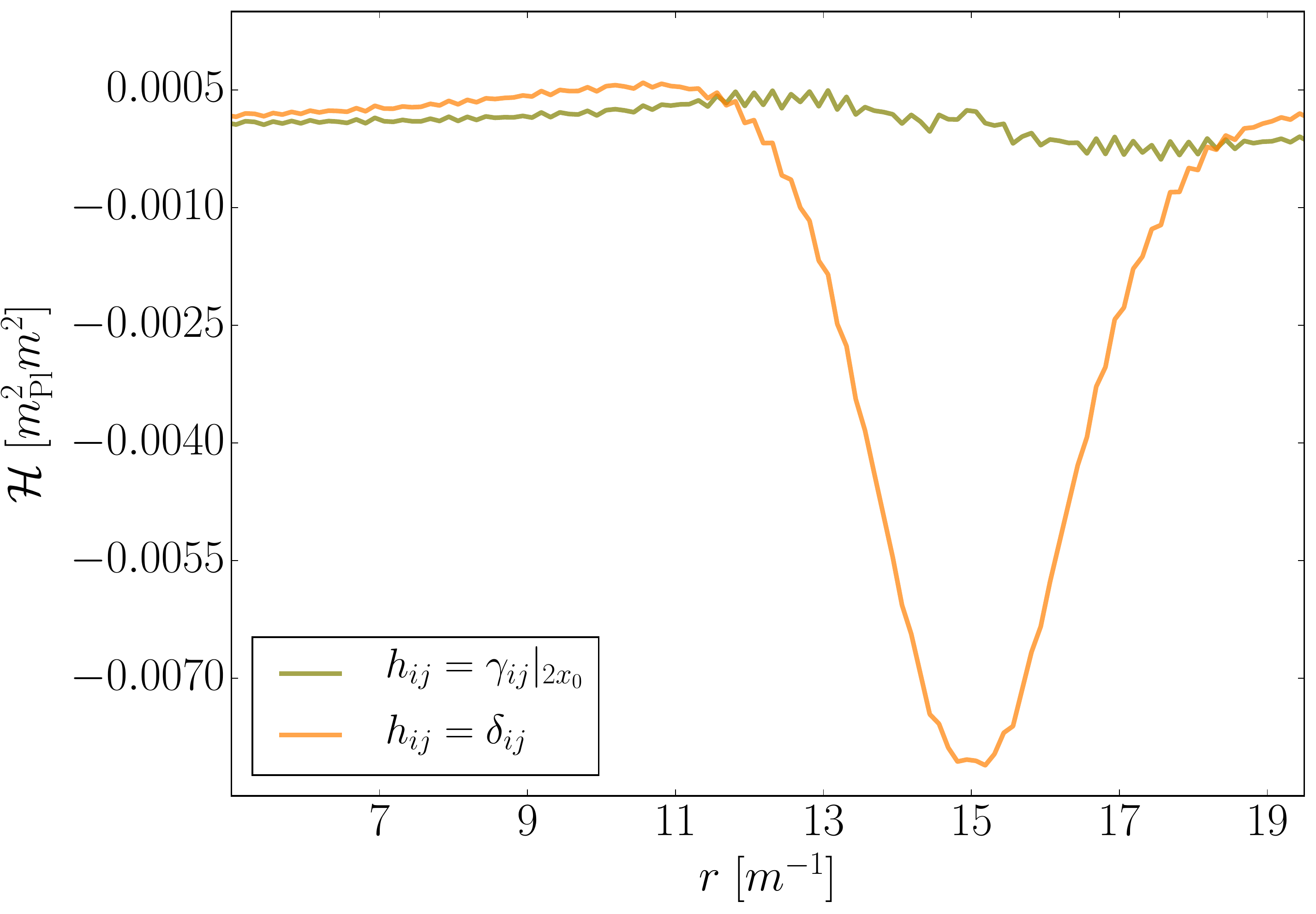}}
\vspace{-1.5em} \caption{
The Hamiltonian constraint violation of the OS-OS initial data before relaxation for $\C = 0.13$  along the axis of the two OS.  By choosing $h_{ij} = \gamma_{ij}|_{2x_0}$ the Hamiltonian constraint violation is reduced by a order of magnitude from 2.6 \% to  0.3 \%. An additional relaxation routine in $\chi$ is implemented after this improvement is applied. }
\label{fig:InitialDataHam}
\end{center}
 \end{figure}
\section{Numerical Methodology}
\subsection{Constructing Initial Data}
\label{sect:Inital_data}

We solve for single oscillaton (OS) profile for $\phi,\pi,\gamma_{ij}$ at some initial hyperslice $t = t_0$ as described in Refs.~\cite{1991PhRvL..66.1659S,Alcubierre:2003sx,UrenaLopez:2002gx,UrenaLopez:2001tw}  where $\pi = \alpha^{-1}(\partial_t\phi - \beta^i\partial_i \phi)$ is the initial kinetic term of the scalar, and $\gamma_{ij}$ is the 3-metric defined as usual in the ADM line element
\begin{equation}
ds^2=-\alpha^2\,dt^2+\gamma_{ij}(dx^i + \beta^i\,dt)(dx^j + \beta^j\,dt).
\end{equation}
The determinant of the spatial metric $\gamma_{\ij}$ will be denoted by $\rm{det}\,\gamma$. We also set the the extrinsic curvature $K_{ij} = 0$. 

\begin{figure}[t!]
\begin{center}
{\includegraphics[width=1.0\columnwidth]{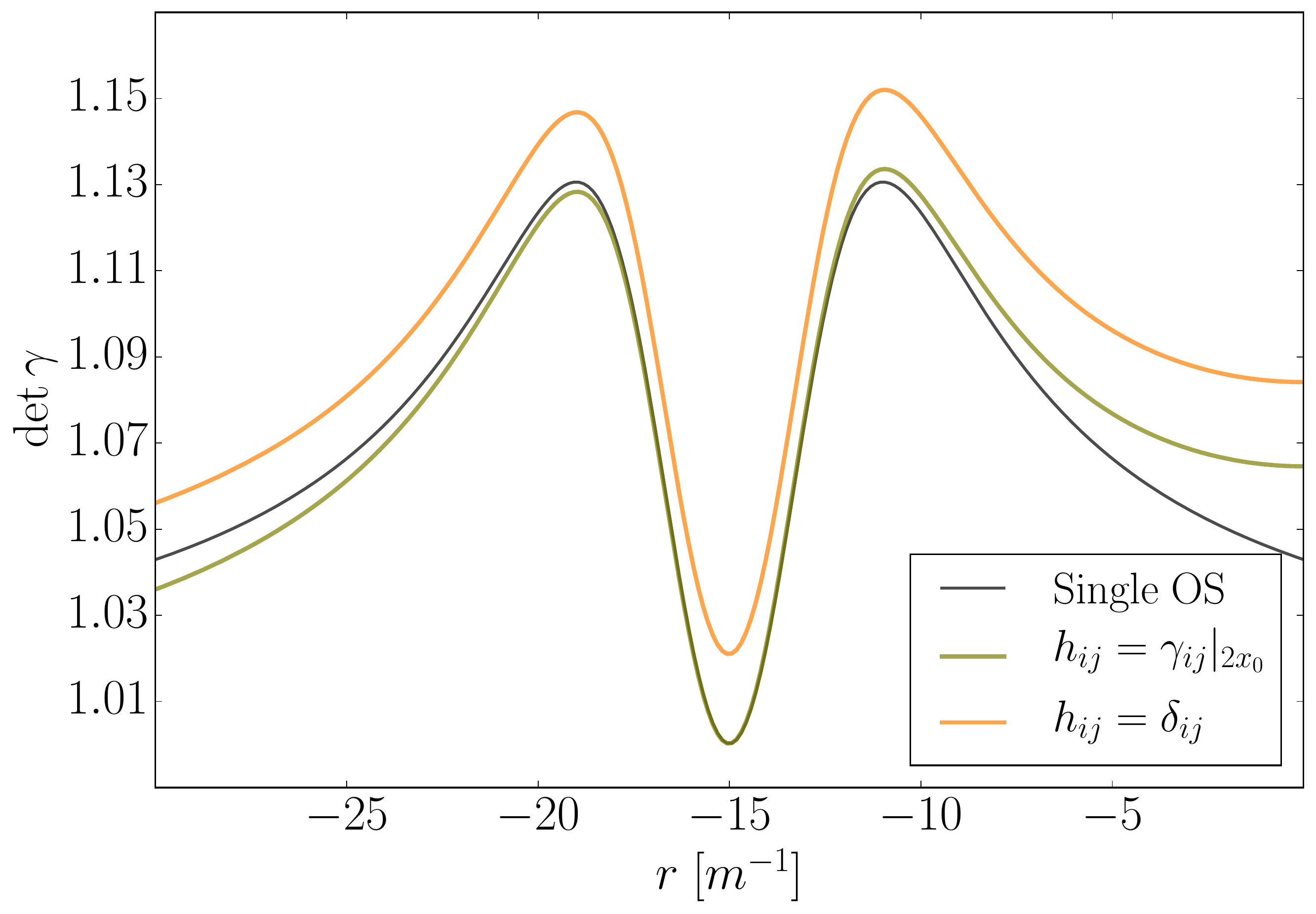}}
\vspace{-1.5em} \caption{
The volume element $\det \gamma$ of the OS-OS initial data before relaxation for $\phi_{m,0}(0) = 0.44$, $\C = 0.13
$ on a line in the $x$ direction which goes through the center of both OS. In this example, we have positioned the stars at distance $\pm  15 ~m^{-1}$ as opposed to $\pm 30~ m^{-1}$ to illustrate our point. Note that the values are closer to the single OS solution when the metric values are conserved in the center of the OS. }
\label{fig:InitialDataDet}
\end{center}
 \end{figure}

Given this single oscillaton profile, we generate static OS-OS initial data by superposing two single OS solutions: 
\begin{equation}
\begin{split}
\phi_{\rm tot} &= \phi|_{x-x_0}+\phi|_{x+x_0}\\
\pi_{\rm tot} &= \pi|_{x+x_0}+\pi|_{x-x_0} \\
\gamma_{ij,{\rm tot}} &= \gamma_{ij}|_{x+x_0}+ \gamma_{ij}|_{x-x_0} - h_{ij} \\
\end{split},
\end{equation}
where $\pm x_0$ are the locations of the centers of the two oscillatons, and $h_{ij}$ is a constant metric. The choice of $h_{ij}$ turns out to be important in setting up the initial conditions. Naively, one would use $h_{ij} = \delta_{ij}$, which would make the asymptotic values the same as for a single OS. However, this choice induces large radial modes in both oscillatons.  These modes are caused by the change in the volume element near the center of each OS   due to the influence of the companion (as compared to the case of an isolated OS). This difference is clearly seen in Fig.~\ref{fig:InitialDataDet}, where the black curve is the volume element related to an isolated OS, whereas the orange curve is the volume element obtained by using $h_{ij}=\delta_{ij}$.
 
An estimate for the change in the volume element can be obtained as follows. Consider ${\rm OS}_1$ at $x_0$, with its companion ${\rm OS}_2$ located at $-x_0$. Assuming a Schwarzchild metric far from the surface of ${\rm OS}_1$, the volume element at $-x_0$ due to ${\rm OS}_1$ is $\sqrt{\det \gamma} = \sqrt{(1-2GM/d)^{-1}}\approx {\cal O}(1.01)$. We assumed a distance $d=2x_0=60 m^{-1}$ and $M\approx 0.5 \mpl^2/m$. By subtracting off $h_{ij}=\delta_{ij}$, we are still left with $\sim 1\%$ extra volume at $x=-x_0$ compared to the case where ${\rm OS}_2$ was isolated (and vice-versa for ${\rm OS}_1$). That is, the oscillatons are ``puffed up'' initially.  These radially excited OS are \emph{not} the initial conditions we seek as they add additional energy and induce instabilities into the initial conditions for single oscillatons.   In particular, the central densities and radii of these excited OS can deviate from their unexcited counterparts by ${\cal O}(100\%)$ and ${\cal O}(10\%)$ respectively as they evolve, potentially rendering any results that we obtain unreliable.\footnote{This is reminiscent of the old ``self-crushing'' problem in the set up of binary neutron stars initial conditions \cite{Suh:2016ctx}.} 
\begin{figure}[t!]
\begin{center}
{\includegraphics[width=1.0\columnwidth]{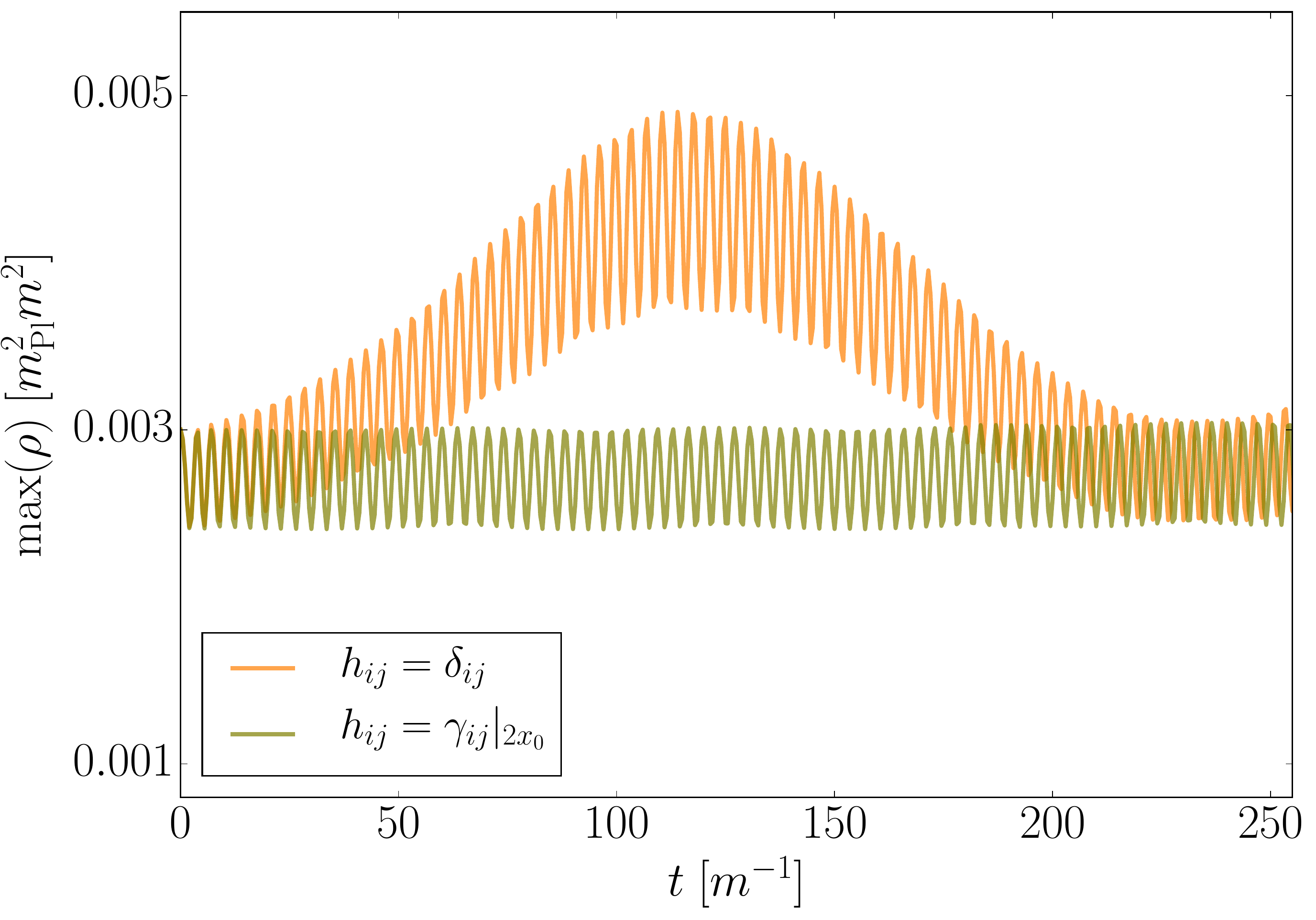}}
\vspace{-1.5em} \caption{
The plot shows the central density of OS with $\C = 0.10 $ at a distance $60m^{-1}$ from its counterpart. The high frequency oscillation with wavelength $\lambda \approx 2\pi m^{-1}$ is the natural breathing of the oscillaton while the low frequency mode is caused by the radial mode. Applying the choice $h_{ij} = \gamma_{ij}|_{2x_0}$ removes this radial mode. Animations depicting the evolution of the central density \underline{\color{blue} \href{https://youtu.be/h0MXVNd8u2E}{with}} and \underline{\color{blue} \href{https://youtu.be/W95UkprupyA}{without}} radial modes are available~\cite{VidwithRadial,VidwithoutRadial} . }
\label{fig:RadialModes}
\end{center}
 \end{figure}
 
As quantitative test, we set up a single OS with compactness $\C = 0.10$, and then imposing a $0.1\%$ perturbation in its radius achieved simply by remapping the field values with $r \rightarrow 0.999r $. This small change results in a large oscillating radial mode with a $\gsim 10 \%$ fluctuation in maximum amplitude of the central density.\footnote{ An \underline{\color{blue} \href{https://youtu.be/LCpTmEuZm-I}{animation}} showing the evolution in time of the central density is available~\cite{VidTest}}  Not surprisingly, the radiated GW energy becomes strongly dependent on the choice of the initial separation causing varying results for different initial distances, thus making it a bad approximation for an unexcited OS falling in from infinity.

Our solution to this problem is to choose $h_{ij} = \gamma_{ij}|_{2x_0}$, which leaves the metric values at the center of each OS unchanged from the isolated case and thus also its volume element.\footnote{A further refinement of this method is to include a factor such that $\lim_{r\rightarrow \infty}  h_{ij} \rightarrow \delta_{ij}$. } From the close match between the green curve ($h_{ij} = \gamma_{ij}|_{2x_0}$) and the black curve (isolated OS) in Fig.~\ref{fig:InitialDataDet}, one can see how this choice is a significant improvement over the orange curve ($h_{ij}=\delta_{ij}$).   

Furthermore, defining the relative Hamiltonian violation as 
\[
	\max\left(\frac{\mathcal{H}}{16\pi\rho}\right),
\]
we see a significant improvement in relative violation from 2.6\% to 0.3\%  (see Fig.~\ref{fig:InitialDataHam}). Finally, we apply a relaxation routine to reduce this Hamiltonian constraint violation further. The result of our method is shown in Fig.~\ref{fig:RadialModes}  where it is clear that we have eliminated the large low frequency radial modes (orange curve) to leave only the physical high frequency central density fluctuation present in the original single OS solution (green curve).

\subsection{Numerical Methodology and Convergence Tests}
\begin{figure}[t!]
\begin{center}
{\includegraphics[width=1.0\columnwidth]{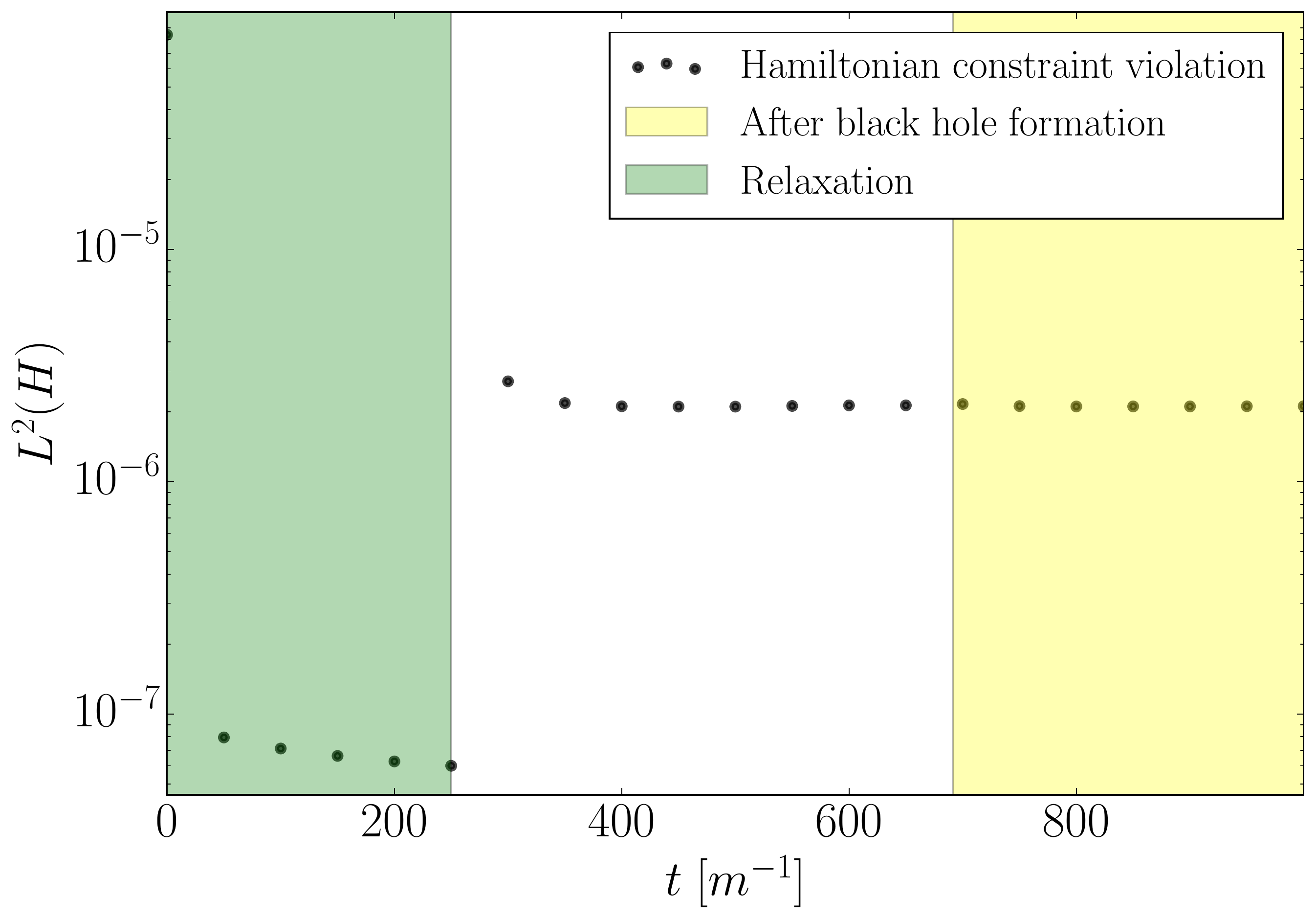}}
\vspace{-1.5em} \caption{
The plot shows the $L^2$ norm (\ref{eq:L2H}) of the Hamiltonian constraint violation over time, with excision of the black-hole interior (which forms around $t=700$). The green region shows the relaxation time, with data points extracted every 100th cycle. There is a jump after the relaxation, likely caused by regridding during transition from relaxation to evolution, but still extremely good overall. } 
\label{fig:L2H}
\end{center}
\end{figure}
All grids for extraction of gravitational waves have a side length of $512 m^{-1}$, with the coarsest resolution being $\Delta x  = 2m^{-1}$. We extract $r\Psi_4$ at a radius of $60m^{-1}$ and we set a fixed resolution over the region containing the extraction sphere. Depending on the scenario, we use from five to six levels of refinement, which corresponds to a smallest resolution of $0.0625m^{-1}$  and $0.03125m^{-1}$, respectively. Since for all simulations the boxsize is $500m^{-1}$  and our extraction sphere is positioned at radius $60m^{-1}$ from the center, we choose the maximum run-time at around $380m^{-1}$ in order to prevent spurious reflections at the boundary from contributing to the final results.

We use the following to measure the volume averaged Hamiltonian constraint violation:
\begin{equation}\label{eq:L2H}
L^2(H) =\sqrt{\frac{1}{V} \int_V |\mathcal{H}^2| dV},
\end{equation}
where $V$ is the box volume with the interior of the apparent horizon excised. As can be seen in Fig. \ref{fig:L2H}, we have good control over the constraint violation throughout the simulation.

\begin{figure}[tb]
\begin{center}
{\includegraphics[width=1.0\columnwidth]{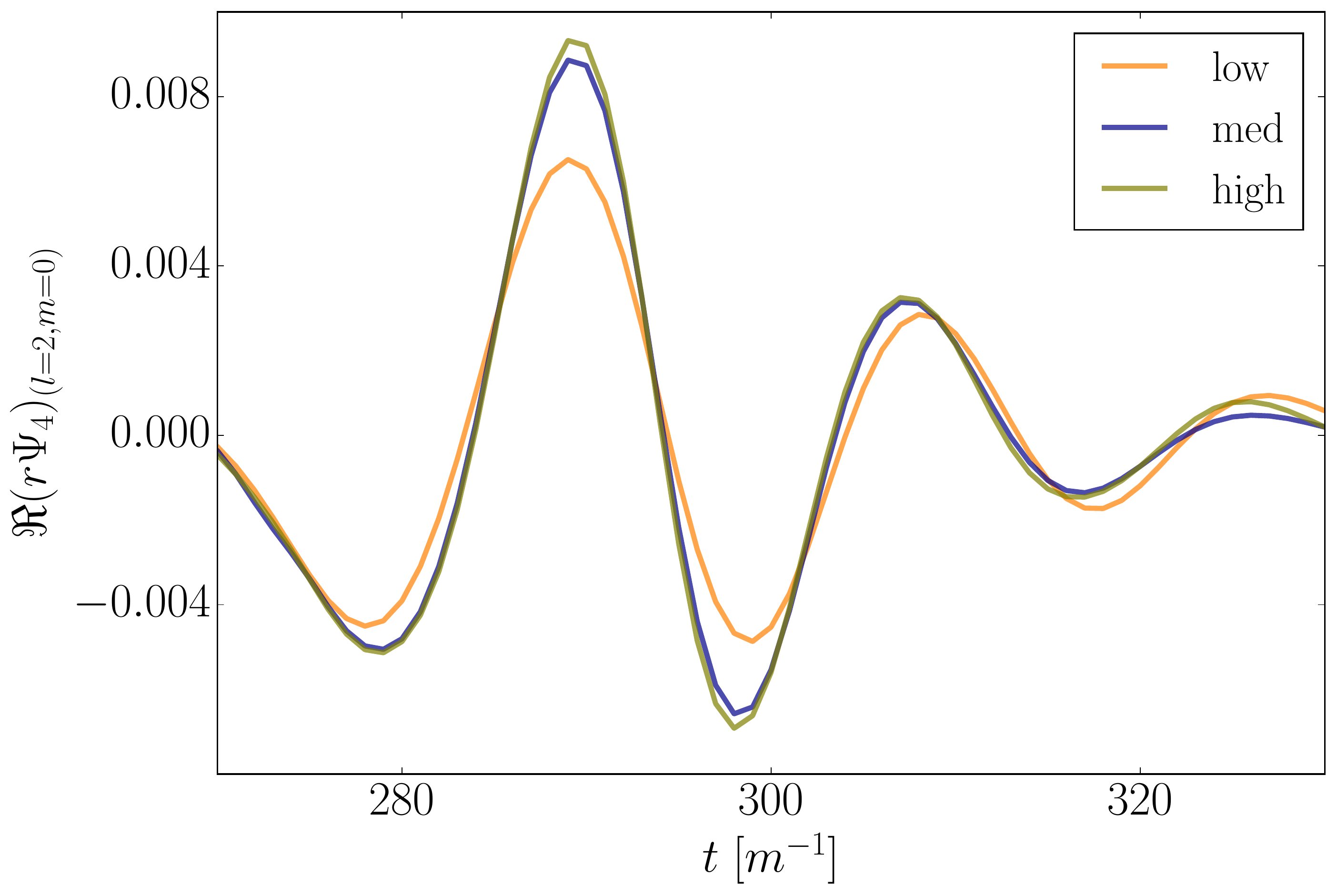}}
\vspace{-1.5em} \caption{Convergence test for the $l = 2$, $m = 0$ mode of $r\Psi_4$, showing convergence between 2nd and 3rd order. The convergence test is done with three different coarsest resolutions of $4m^{-1}$, $2m^{-2}$ and $1m^{-1}$, 6 levels of $2:1$ refinement, with corresponding finest resolutions of $0.0625m^{-1}$, $(0.0625/2)m^{-1}$ and $(0.0625/4)m^{-1}$. Our evolution scheme is 4th order, and the lowered accuracy is due to the large amount of re-gridding required to track the motion of the oscillaton through to final state. 
}
\label{fig:Connvergence}
\end{center}
\end{figure}

We test the convergence of our simulations with the collision of two oscillatons with $\phi_{m,0}(0)= 0.33$ ($\mathcal{C}$= 0.20).  We use a box of  sidelength $256m^{-1}$, and initial separation of the oscillatons of $40 m^{-1}$. As we have turned on adaptive refinement, we use three different coarse resolutions of $1m^{-1}$, $2m^{-1}$ and $4m^{-1}$. This allows for 6 levels of $2:1$ refinement each  with corresponding finest possible resolutions of $0.015625m^{-1}$, $0.03125m^{-1}$ and $0.0625m^{-1}$. We extract the $l = 2$, $m = 0$ mode of $r\Psi_4$ at distance $60m^{-1}$ from the center. The results are shown in Fig. \ref{fig:Connvergence}, where we obtain  between 2nd and 3rd order convergence on average. While we have used a 4th order scheme, the large amount of re-gridding required to track the collision results in some loss of accuracy which is not surprising.\footnote{Using fixed grids, we have demonstrated 4th order convergence of the code consistent with methods used \cite{Clough:2015sqa,Helfer:2016ljl}.}
Lastly, we note that an estimate for the error bars in the energy extraction (Fig.~\ref{fig:MoneyPlot}
) is obtained by doubling the resolution of the simulations described in the main text, and computing the energy for this higher resolution case. The difference of the results at two different resolutions gives us an estimate for the error. 

\section{Self-Interactions}
In our study we have ignored possible self-interactions of the scalar field $\phi$. Here we discuss the domain of validity of our results. 

Let us first consider the case where our compact scalar solitons are made of axionic dark matter. In this case, the potential $V(\phi)=m^2f^2[1-\cos\phi/f]=m^2\phi^2/2-\lambda\phi^4/4!+\hdots$, where $\lambda=m^2/f^2$. By comparing the self-interaction and the gravitational interaction, the gravitational interaction dominates our solitons for $\phi/f\lesssim \mathcal{C}^{1/2}$ (where the dimensionless compactness $\mathcal{C}=GM/R$ is of the order of the typical gravitational potential associated with each soliton).\footnote{For non-axionic cases with an attractive self-interaction:  $\phi\lesssim (m/\sqrt{\lambda}) \mathcal{C}^{1/2}$.} For our merger simulations, the maximum value of the field is typically $\phi_{\rm max}\lesssim 0.24\,m_{\rm Pl}$ (estimated as twice the maximum field value at the center of individual oscillatons). Hence, for $f\gtrsim m_{\rm pl}$, we expect our results will remain unchanged. 

Although it is not impossible to envision a mechanism through which such a large value of the decay constant would be set in the effective theory~\cite{Kim:2004rp,Kappl:2014lra}, $f\gtrsim m_{\rm Pl}$ is phenomenologically problematic if $\phi$ constitutes the totality of dark matter. Assuming we have a scenario similar to \cite{Hui:2016ltb}, for $f\gtrsim m_{\rm Pl}$, the total dark matter abundance bound requires the axion to be  unacceptably light ($m< 10^{-30}\,{\rm eV}$), in conflict with observations~\cite{Marsh:2015xka}. An obvious way around this abundance bound is to assume that the field $\phi$ corresponds to a sub-dominant dark matter component. Conservatively, the results of our paper are therefore expected to apply for solitons made of a subdominant axionic dark matter component with $f\gtrsim m_{\rm Pl}$.

As we have discussed, the upper bound of $f\lesssim m_{\rm Pl}$ is desirable from both a model building perspective and from abundance constraints. 
In the regime $f\ll m_{\rm Pl}$, we would expect self-interactions to be relevant.

However, for $f\lesssim 10^{-2} m_{\rm Pl}$, the self-gravitating real-scalar lumps cannot reach compactness values that would make them approximate mimickers of BHs~\cite{Helfer:2016ljl}. Hence, including self-interactions, a typical $f$ value of interest for gravitational wave emission would be $f\sim 10^{-1}m_{\rm Pl}$. 
For this value of $f$, we have found that the compactness of our configuration of mass $M$ can change by {\it at most $20\%$} relative to the non-interacting case. How does this affect our results?  While the compactness for a given mass changes, {\it if} the fractional gravitational wave output is a function of compactness only, our curve in Fig.~1 should remain unchanged. 

These heuristic arguments deserve a more complete study, which will be taken up in future works.
The main difficulty lies in setting up initial conditions. The construction of an unexcited ultra-compact initial configuration with significant self-interactions within full nonlinear GR is still an open problem~\cite{Helfer:2016ljl}. 

\bibliography{mybib,axion_star,Refs}

\clearpage
\appendix

\end{document}